\documentstyle[11pt,psfig,amssymb]{article}
\newcommand{\RR}{\rangle \rangle}
\newcommand{\LL}{\langle \langle}

\begin{document}
\renewcommand{\thepage}{ }
\begin{titlepage}
\title{
{\center \bf N\'eel ordered {\sl versus} quantum disordered
behavior in doped spin-Peierls and Haldane gap systems}}
\author{
R. M\'elin\thanks{melin@polycnrs-gre.fr}
{}\\
{}\\
{Centre de Recherches sur les Tr\`es Basses
Temp\'eratures (CRTBT)\thanks{U.P.R. 5001 du CNRS,
Laboratoire conventionn\'e avec l'Universit\'e Joseph Fourier
}}\\
{CNRS BP 166X, 38042 Grenoble Cedex, France}\\
{}\\
{Laboratoire de Physique\thanks{U.M.R. CNRS 5672},
Ecole Normale Sup\'erieure de Lyon}\\
{46 All\'ee d'Italie, 69364 Lyon Cedex 07, France}\\
{}\\
}
\date{\today}
\maketitle
\begin{abstract}
\normalsize
I consider a theoretical description of recent
experiments on doping the spin-Peierls
compound CuGeO$_3$
and the Haldane gap compounds PbNi$_2$V$_2$O$_8$ and Y$_2$BaNiO$_5$.
The effective theory is the one of randomly distributed
spin-$1/2$ moments interacting with an exchange
decaying exponentially with distance.
The model has two phases in the (doping, interchain
coupling) plane: (i) A N\'eel ordered phase at small
doping; (ii) A quantum disordered phase at larger
doping and small interchain interactions. The
spin-Peierls compound CuGeO$_3$ and the
Haldane gap Nickel oxides PbNi$_2$V$_2$O$_8$ and Y$_2$BaNiO$_5$ fit well into this phase
diagram. At small temperature, the N\'eel
phase is found to be reentrant into the quantum disordered
region. The N\'eel transition 
relevant for CuGeO$_3$ and PbNi$_2$V$_2$O$_8$ can
be described in terms of a classical disordered
model. A simplified version of this model is introduced,
and is solved on a hierarchical lattice structure,
which allows to discuss the renormalization group flow
of the model.
It is found that the system looks non disordered at large scale,
which is not against available susceptibility 
experiments.
In the quantum disordered regime
relevant for Y$_2$BaNiO$_5$, the two spin model
and the cluster RG in the 1D regime show a 
power law susceptibility, in agreement with
recent experiments on Y$_2$BaNiO$_5$. It is found that
there is a
succession of two distinct quantum disordered phases
as the temperature is decreased.
The classical disordered model of the doped spin-$1$
chain
contains already
a physics relevant to the quantum disordered phase.
\end{abstract}
\end{titlepage}
\newpage
\renewcommand{\thepage}{\arabic{page}}
\setcounter{page}{1}
\baselineskip=17pt plus 0.2pt minus 0.1pt

\newpage




\section{Introduction}

Doping quasi one dimensional (1D) antiferromagnets
with a spin gap
has become experimentally possible since the
discovery of several inorganic quasi 1D oxides.
One of these compounds is CuGeO$_3$ having a
spin-Peierls transition at $T_{SP} \simeq 14$~K~\cite{SP}.
Below $T_{SP}$, the spin-phonon coupling induces
a dimerization of the lattice, and the opening of
a gap in the spin excitation spectrum.
The Haldane gap in spin-$1$ chains in another example
of a spin gap state in low dimensional magnets~\cite{Haldane}.
Two inorganic
spin-$1$ Haldane gap antiferromagnets have
been discovered in the recent years:
(i) PbNi$_2$V$_2$O$_8$ having a spin gap $\simeq 28$~K~\cite{Pb}.
(ii) Y$_2$BaNiO$_5$ having a spin gap $\simeq 100$~K~\cite{ori-Y}.
The spin-Peierls compound CuGeO$_3$ and the two
Nickel oxides PbNi$_2$V$_2$O$_8$ and Y$_2$BaNiO$_5$ can 
be doped in a very controlled fashion.
Substituting the magnetic Cu sites (having $S=1/2$)
of the spin-Peierls compound CuGeO$_3$
with a variety
of ions (Ni~\cite{Ni} -- a spin-1 ion --,
Co~\cite{Co} -- a spin-$3/2$ ion --,
Zn~\cite{Zn,Hase,Martin,Grenier,Saint-Paul} or Mg~\cite{Mg}
-- non magnetic ions --),
or substituting the Ge sites with Si~\cite{Si}
leads to the formation of an antiferromagnetic
phase (AF) at low temperature. Moreover,
in CuGeO$_3$, there is AF long range
order even with an extremely weak concentration
of Zn impurities~\cite{Manabe98}.
On the other hand, the two Nickel oxides
PbNi$_2$V$_2$O$_8$ and Y$_2$BaNiO$_5$ have been the subject of an important
experimental interest recently.
It has been shown
that substituting the spin-$1$ Ni sites of the PbNi$_2$V$_2$O$_8$
compound with Mg -- a spin-$0$ ion -- leads
to AF long
range order. In the Y$_2$BaNiO$_5$ compound,
the Ni sites can be substituted with
Zn or Mg -- non magnetic ions --. In this case, no
sign of AF long range order has been reported,
even at extremely low
temperature~\cite{Batlogg,DiTusa,Kojima}. Instead,
it has been found that the susceptibility
has a power-law temperature dependence~\cite{Payen00}.
Experiments therefore show that doping quasi 1D
antiferromagnets with a spin gap can lead to 
very different situations: either antiferromagnetism,
or a power-law susceptibility without AF ordering.
The purpose
of the present article is to
describe these experimental observations in a unified
theoretical framework,
and provide a detailed theoretical analysis of the
different phases of the model.

On the theoretical side, a lot of efforts have been
devoted to understand the behavior of random spin chains.
The theoretical tool usually used to study
these models is the cluster renormalization group (RG)~\cite{Dasgupta},
which is a perturbation theory in the inverse of the
strength of the strongest exchange in the chain,
and leads to a certain number of exact results at
low temperature
because the exchange distribution becomes extremely
broad. A lot of different models have been solved
using this approach. For instance: the Ising chain
in a transverse magnetic field~\cite{Fisher92},
the random spin-$1/2$ chain~\cite{Fisher94},
the spin-$1/2$ chain with random ferromagnetic
and antiferromagnetic bonds~\cite{Westerberg},
the dimerized chain with random bonds~\cite{Girvin96},
the disordered Haldane gap chain~\cite{Hyman,Monthus}.
These studies have revealed that 1D disordered
magnets can be controlled by several types
of Griffiths phases: (i) the
random singlet phase with a diverging susceptibility
and algebraic correlation; (ii) the ``weakly disordered''
phase with a diverging susceptibility and
short range correlations.

An important question is to understand the relation between
the available experiments on quasi one dimensional
oxides and the available theories of disordered
1D magnets. This type of approach followed
in the present article
incorporates realistic constraints such as
interchain couplings and a finite temperature.

The starting point of such our description
has been already established in previous
works~\cite{Fabrizio97a,Fabrizio97b,Fabrizio99,Melin00,Dobry99}.
In the doped spin-Peierls systems, non
magnetic impurities generate solitonic
spin-$1/2$ degrees of freedom distributed
at random with a concentration $x$.
The solitons are confined
close to the impurities because of interchain
interactions~\cite{Dobry99,Khomskii,Fuku,Mosto,Laukamp98,Hansen98a,Hansen98b,Augier98},
and
interact with the Hamiltonian
\begin{equation}
\label{eq:H}
{\cal H} = \sum_{\langle i,j \rangle}
J_{i,j} {\bf S}_i . {\bf S}_j
.
\end{equation}
The exchange between two spin-$1/2$ moments at 
positions $(x_i,y_i)$ and $(x_j,y_j)$ is mediated
by virtual excitations of the gaped medium and
therefore decays exponentially with distance:
\begin{equation}
J_{i,j} = (-)^{x_i - x_j + y_i - y_j}
\Delta \exp{\left( - \sqrt{
\left( \frac{x_i - x_j}{\xi_x} \right)^2 +
\left( \frac{y_i - y_j}{\xi_y} \right)^2}\right)}
\label{eq:J}
,
\end{equation}
where $\xi_x$ and $\xi_y$ are the correlation lengths
in the direction of the chains and perpendicular
to the chains respectively~\cite{Fabrizio99,Melin00,Dobry99}.
The form Eq.~\ref{eq:J}
of the exchange incorporates a correlation length
in the transverse direction shorter than in the
longitudinal direction ($\xi_y = \xi_x /10$,
and $\xi_x \simeq 10$ in CuGeO$_3$~\cite{Kiryukin,Horvatic}).
The exchange Eq.~\ref{eq:J}
is staggered because the dimerized pattern
propagates staggered antiferromagnetic correlations.
The Hamiltonian Eqs.~\ref{eq:H},~\ref{eq:J} is therefore
strongly disordered but unfrustrated.

\begin{figure}
\centerline{\psfig{file=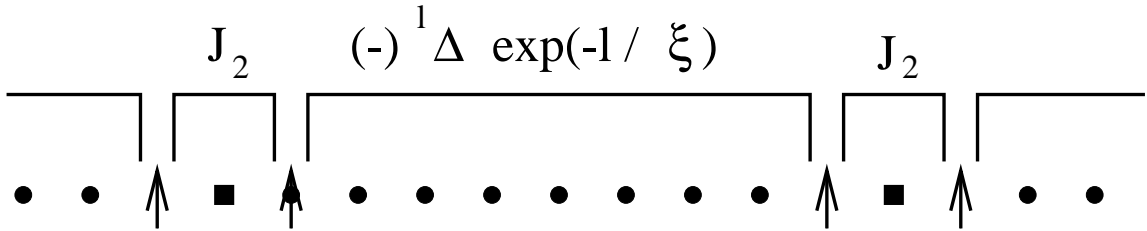,height=3cm}}
\caption{The low energy effective model of the
doped spin-$1$ system.
A paramagnetic
impurity (square symbols) generates a unit
of two edge moments.
The edge moments originating from the same
impurity are coupled ferromagnetically by an
exchange $-J_2$ having the same order of magnitude
as the interchain coupling.
Two edge moment at a distance $l$ are coupled
by the exchange $(-)^l \Delta \exp{(-l/\xi)}$.
}
\label{fig:schema}
\end{figure}

Now, the model relevant to describe doping in
a Haldane gap system is almost identical. It is well known
that an impurity in a Haldane gap
chain generates two ``edge'' spin-$1/2$ moments:
one at the right and one at the left of the impurity
site~\cite{Kennedy,Hagiwara90,Sorensen}.
The spin-1 chain can be thought in terms
of a Valence Bond Solid (VBS)~\cite{AKLT}. 
Introducing a paramagnetic site breaks two
VBS bonds, therefore resulting in two ``edge''
spin-$1/2$ moments. At energies far below the
Haldane gap, only these edge moments
are the relevant degrees of freedom (see
Fig.~\ref{fig:schema}).
The two edge moments in the same unit
({\sl i.e.} generated by the same impurity)
interact with a ferromagnetic exchange $-J_2$ originating from
the coupling to neighboring chains, with therefore 
the same order of magnitude as the interchain
interaction: $J_2 \sim J_{\perp}$~\cite{Fabrizio-spin1}.
The edge moments belonging to different units
are coupled by the staggered exchange
Eq.~\ref{eq:J}.

We find that, depending on the doping concentration
and interchain interactions, the model Eqs.~\ref{eq:H},~\ref{eq:J}
has two regimes: a N\'eel ordered region and a quantum disordered
region.
In the N\'eel ordered
region, relevant for CuGeO$_3$ and PbNi$_2$V$_2$O$_8$,
quantum mechanics plays little role,
and we are lead to replace the spin variables
in Eq.~\ref{eq:H},~\ref{eq:J} by classical
Ising spins.
We propose here that this type of disordered
Ising model is equivalent to another type of disordered
Ising model, and solve the hierarchical lattice version
of the latter. Using this treatment, we can compute
the renormalized exchange distribution.
In spite of a strongly
disordered initial Hamiltonian (see Eqs.~\ref{eq:H},~\ref{eq:J}),
we find that at large scale, the problem behaves as if it
were non disordered. This appears to be consistent with
susceptibility experiments showing a well-defined
transition even with a small doping concentration~\cite{Manabe98}.

In the quantum disordered region of the phase
diagram, the physics is dominated by the formation
of randomly distributed singlets.
We show that the susceptibility has a power-law
behavior, which turns out to be in agreement
with existing experiments on Y$_2$BaNiO$_5$~\cite{Payen00}. We also find
the existence of two distinct ``quantum disordered''
phases. The high temperature ``quantum disordered'' phase
appears to have been observed experimentally in
Y$_2$BaNiO$_5$~\cite{Payen00}. There appears to be another low
temperature ``quantum disordered'' phase in which
the edge spins in the same unit (see Fig.~\ref{fig:schema})
are frozen into spin-$1$ objects.
Finally, we
show that even in the quantum disordered regime of the model,
part of the quantum disordered behavior is already contained
in the classical disordered magnet.

The article is organized as follows: the phases
of the model Eq.~\ref{eq:H},~\ref{eq:J} are discussed in
section~\ref{sec:phase-dia}. We show the existence
of two phases: a N\'eel ordered region and a quantum disordered
region. The nature of these two phases is next
discussed in details in sections~\ref{sec:classical}
and~\ref{sec:nature-qu}. Concluding remarks are given
in section~\ref{sec:conclusion}.

\section{Phases of the model}
\label{sec:phase-dia}
In this section, we use a phenomenological approach
to derive the phase diagram of the model.
The calculation of the relevant energy scales
in the problem is based on the
analysis of a two-spin model.
There is a first temperature scale (being a fraction
of the spin gap $\Delta$)
below which magnetic
correlations start to develop inside the chains.
There is a second energy scale $T_{\rm typ}$,
equal to the typical exchange,
associated to singlet formation.
There is a third energy scale
$T_{\rm Stoner}$
associated to long range AF ordering.
The behavior of the model depends strongly
on whether $T_{\rm Stoner}$ is larger or smaller
than $T_{\rm typ}$. 

\subsection{Onset of magnetic correlations}
\label{sec:Jav}
\begin{figure}
\centerline{\psfig{file=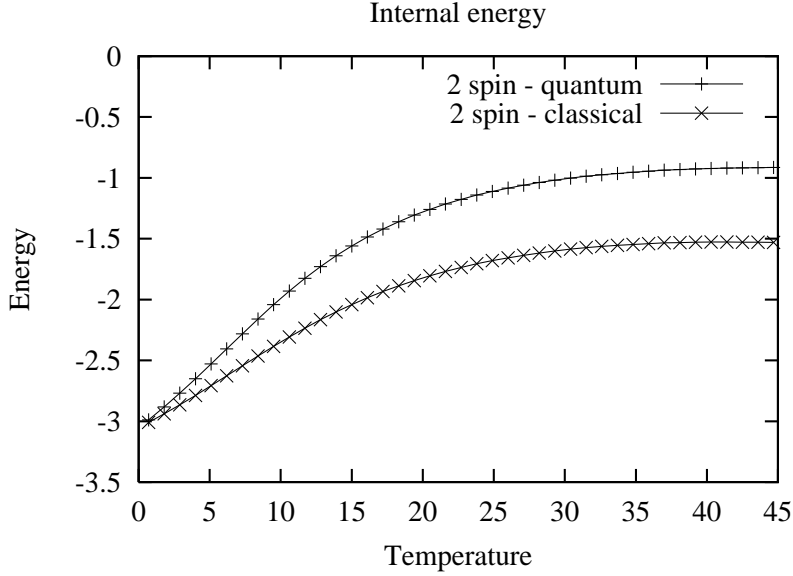,height=8cm}}
\caption{Temperature dependence of the internal
energy of the classical and quantum two-spin model,
with $\Delta=44.7$~K, $\xi=10$,
$x=0.01$, relevant for CuGeO$_3$.
The temperature and energy are calculated in Kelvin. 
In the two-spin classical model, we considered two spins
coupled by an Ising term ${\cal H} = (J/2)
\sigma_1 \sigma_2$.
}
\label{fig:U}
\end{figure}

Magnetic correlations start to appear inside the
chains when the temperature is a fraction of the
spin gap $\Delta$.
To show this, let us consider a simple model in which
two spins at a distance $l$ are
coupled antiferromagnetically:
${\cal H} = J(l) {\bf S}_1 . {\bf S}_2$. We use
an exchange decaying exponentially with distance
(see Eq.~\ref{eq:J}):
$J(l) = \Delta \exp{(-l / \xi)}$, and a Poisson bond
length distribution ${\cal P}(l) = x \exp{(-x l)}$. 
Rigorously, the spacing $l$ is a discrete quantity,
distributed according to a geometrical distribution.
However, the physics will turn out to be controlled
by the large-$l$ behavior and it is legitimate
to consider $l$ as a continuous variable,
and replace the geometrical distribution by the 
Poisson distribution.

The internal energy of the two-spin model reads
\begin{equation}
\label{eq:U-RS}
U(T) = - \frac{3}{4} x \xi \Delta^{-x \xi}
T^{x \xi +1} \int_0^{\beta \Delta} 
u^{x \xi} \frac{ \exp{(3 u/4)} -  \exp{(-u/4)}}
{\exp{(3 u/4)} + 3 \exp{(-u/4)}} du
.
\end{equation}
This expression can be expanded
in $T$:
\begin{equation}
\label{eq:U-0-RS}
U(T) \simeq - \frac{3}{4} \frac{ x \xi \Delta}
{1 + x \xi} + 
\frac{9}{2} x \xi \left( \frac{T}{\Delta} \right)^{x \xi}
T + ...
\end{equation}
Magnetic correlations start to appear in the two-spin
model in the low temperature regime in which
$U(T)$ is linear in $T$. This regime appears when $T$ is
a fraction of $\Delta$ (see Fig.~\ref{fig:U}).

\subsection{Singlet formation}
\label{sec:singlet-for}
To discuss in what temperature range is the physics
controlled by the quantum mechanical ground
state (being a singlet), we need to calculate
the probability ${\cal P}_s(T)$
to find the two spins
in a singlet state at a 
finite temperature $T=1/\beta$. We have
\begin{eqnarray}
{\cal P}_s(T) &=& \int dl {\cal P}(l) \frac{ \exp{ [3 \beta J(l) / 4]}}
{\exp{[3 \beta J(l) / 4]} + 3 \exp{[-  \beta J(l) / 4]}}\\
&=& 1 - 3 x \xi \left(
\frac{T}{\Delta} \right)^{x \xi} \int_0^{\beta \Delta}
u^{-1 + x \xi} \frac{ \exp{(-u/4)} }
{ \exp{(3 u /4)} + 3 \exp{ (-u/4)}} du
\label{eq:Psbis}
,
\end{eqnarray}
where we used the dimensionless parameter
$u= \beta J$. The integral in Eq.~\ref{eq:Psbis} 
is dominated by the small exchanges and we have
${\cal P}_s(T) \simeq
1 - \kappa (T/\Delta)^{x \xi}$, with $\kappa$
a numerical factor. We are lead to conclude that
the ground state occupancy is close to unity below
the energy scale $T_{\rm typ} \sim \Delta
\exp{(-1/(x \xi))}$. $T_{\rm typ}$ is nothing
but the typical exchange, already identified in
a previous work on the 1D
model~\cite{Fabrizio97a,Fabrizio97b}.

The calculation of the ground state occupancy
can be made
even simpler by noticing that only the disorder
configurations in which the exchange is larger
than $\sim T$ are in a singlet configuration.
This leads to
$$
{\cal P}_s(T) \simeq \int_0^{\xi \ln{(\beta \Delta)}}
x \exp{(-x l)} dl \simeq 1 -
\left(\frac{T}{\Delta}\right)^{x \xi}
.
$$
It is remarkable that the energy scale $T_{\rm typ}$
arising from the two-spin model is {\sl identical}
to the one obtained previously from the exact
solution of the 1D effective
model of the
spin-Peierls chain~\cite{Fabrizio97b}. 
This shows that a model with only two spins contains
already the relevant physics.

\begin{figure}
\centerline{\psfig{file=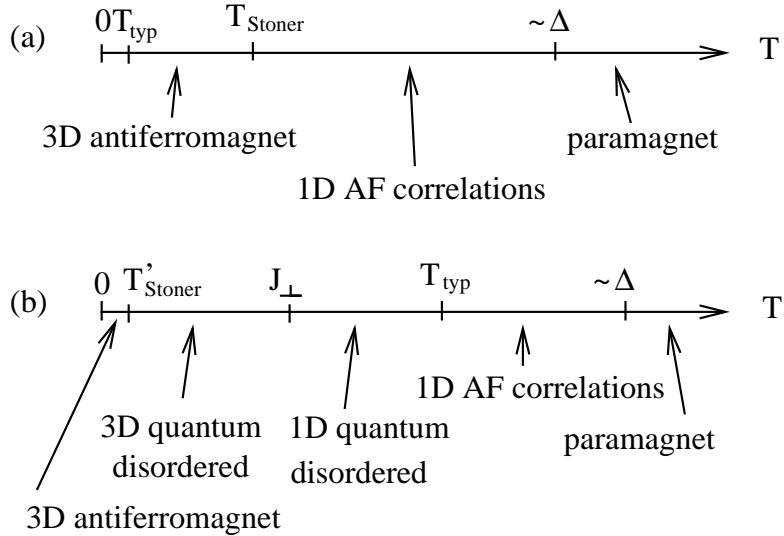,height=7cm}}
\caption{In the models with 3D antiferromagnetism (a)
$T_{\rm typ}$ is far below $T_{\rm Stoner} =
J_{\perp} x \xi$, with therefore a classical
transition to an AF phase. This is the case
for PbNi$_2$V$_2$O$_8$ as well as CuGeO$_3$.
In the models with a quantum disordered ground state (b),
$J_{\perp}$ is far below $T_{\rm typ}$,
which occurs in Y$_2$BaNiO$_5$.
In this
case, the classical paramagnet has a cross-over to
a 1D quantum disordered state with formation
of random singlets. Below $J_{\perp}$,
singlets are formed among spins belonging to different
chains. Below $T_{\rm Stoner}'$, there is a transition
to a reentrant antiferromagnetic phase. Intrachain
correlations start to play a role when the temperature
is below a fraction of the spin gap $\Delta$.
}
\label{fig:regime}
\end{figure}

\subsection{N\'eel ordered {\sl versus} quantum disordered
behavior}
\label{sec:Neel-order}
At low temperature, correlations between the chains
induce a long range ordering
of the spin system. The simplest phenomenological
description of long
range ordering is provided by a Stoner model,
already considered in Ref.~\cite{Fabrizio99}.
In CuGeO$_3$, there is a succession of three regimes
(see Fig.~\ref{fig:regime}-(a)):
(i) a paramagnet at high temperature,
(ii) intrachain correlations develop when the temperature
is a fraction of $\Delta$ (iii)
interchain correlations give rise to
long range antiferromagnetism below $T_{\rm Stoner}
= J_{\perp} x \xi$.

In PbNi$_2$V$_2$O$_8$, the relevant parameters are
$\Delta \simeq 30$~K, $x \simeq 0.02$,
$J_{\perp} \simeq 1.1$~K. We approximate the correlation
length in the Haldane gap phase to be $\xi \simeq 6$.
The true correlation length is expected to be larger
than this value because PbNi$_2$V$_2$O$_8$ is close
to a transition to
an Ising ordered antiferromagnet~\cite{Pb}.
We find
$T_{\rm Stoner}= 0.13$~K, $T_{\rm typ} = 6$~mK,
showing that the same succession of regimes
occur in PbNi$_2$V$_2$O$_8$ and CuGeO$_3$ (see Fig.~\ref{fig:regime}-(a)).
This is  compatible with existing experiments
in PbNi$_2$V$_2$O$_8$~\cite{Pb}.

Therefore, in CuGeO$_3$ and PbNi$_2$V$_2$O$_8$, one has $T_{\rm Stoner}
\gg T_{\rm typ}$. This implies that singlet formation
plays little role in the physics of the antiferromagnetic
transition. The
quantum two-spin model can then be well
mimicked by the {\sl classical}
two-spin model. For instance, the internal
energy of the classical and quantum two-spin models
have an identical temperature dependence
(see Fig.~\ref{fig:U}). This indicates that
one might expect to obtain a reasonable description
of antiferromagnetism in CuGeO$_3$ and PbNi$_2$V$_2$O$_8$ on the basis
of a classical model, which we analyze in details in
section~\ref{sec:classical}.

Now, the situation is different in Y$_2$BaNiO$_5$, where one has
$\Delta \simeq 100$~K, $\xi \simeq 6$,
$J_{\perp} \simeq 0.3$~K, and $x \simeq 0.04$~\cite{Payen00}.
We find $T_{\rm typ}=1.6$~K, and
$T_{\rm Stoner} = 0.07$~K.  What is new compared
to CuGeO$_3$ and PbNi$_2$V$_2$O$_8$ is that
a well defined quantum disordered regime is present
below $T_{\rm typ}$.
In between $J_{\perp}$ and $T_{\rm typ}$, there is
singlet formation in the chain direction, and below
$J_{\perp}$, the singlets develop in
the transverse direction. Below the energy scale $T_{\rm typ}$,
the staggered susceptibility is well
described by the one of the random spin-$1$ chain~\cite{Hyman,Monthus}
\begin{equation}
\label{eq:chi-power}
\chi(T) \sim x T^{\alpha-1} / T_{\rm typ}^\alpha
.
\end{equation}
Using a cluster RG calculation, 
we will calculate the susceptibility in section~\ref{sec:clusterRG}
and show that it has indeed a power-law temperature
dependence. The Stoner criterion leads to the ordering
temperature
$$
\frac{T_{\rm Stoner}'}{T_{\rm typ}} =
\left( \frac{ J_{\perp} x \xi}{T_{\rm typ}}
\right)^{1/(1-\alpha)}
.
$$
This shows that the quantum disordered
model transits to a reentrant antiferromagnetic ground
state below $T_{\rm Stoner}'$.  The
different regimes of the model are
shown on Fig.~\ref{fig:regime}-(b).

\begin{figure}
\centerline{\psfig{file=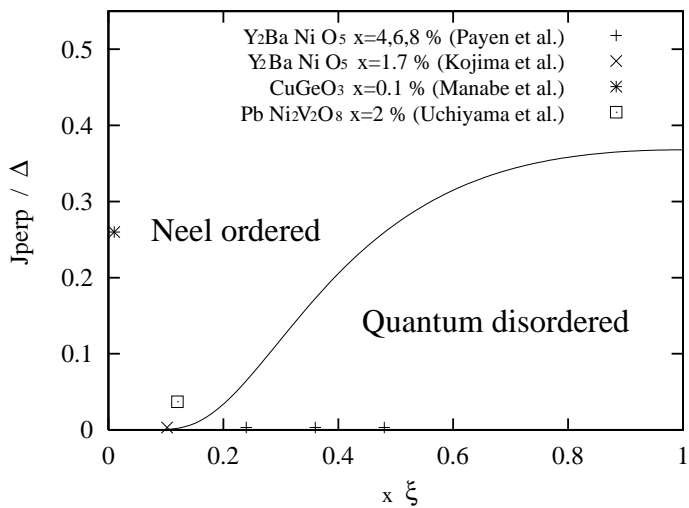,height=7cm}}
\caption{Phase diagram in the plane $(x \xi,
J_{\perp}/\Delta)$. The solid line is
$J_{\perp}/\Delta=1/(x \xi) \exp{[-1/(x \xi)]}$,
delimitating the transition from the N\'eel ordered
region to the quantum disordered region. Various
experimental systems have been reported on the diagram:
$+$, $\times$: Y$_2$BaNiO$_5$~\cite{Payen00,Kojima}
(being quantum disordered);
*: CuGeO$_3$~\cite{Manabe98} (being antiferromagnetic);
$\Box$: PbNi$_2$V$_2$O$_8$~\cite{Pb} (being antiferromagnetic).
The values of $J_{\perp}$
and $\xi$ have been taken from these references.
This phase diagram is valid above the temperature
scale $T_{\rm Stoner}'$. Below $T_{\rm Stoner}'$, the
antiferromagnet is reentrant inside the quantum disordered
phase.
}
\label{fig:diagram}
\end{figure}

The existence of two classes of models is best summarized
by calculating the ratio
$$
\frac{T_{\rm Stoner}}{T_{\rm typ}} = \frac{J_{\perp}}{\Delta}
x \xi e^{1/(x \xi)}
,
$$
which can be smaller
or larger than unity, controlling whether the model
has a N\'eel transition at $T_{\rm Stoner}$
or is in a quantum disordered regime. These two behaviors,
as well as a comparison between the model and existing
experiments on CuGeO$_3$, PbNi$_2$V$_2$O$_8$ and Y$_2$BaNiO$_5$, have been reported on the
phase diagram on Fig.~\ref{fig:diagram}. 

\section{Nature of the antiferromagnetic transition}
\label{sec:classical}
\begin{figure}
\centerline{\psfig{file=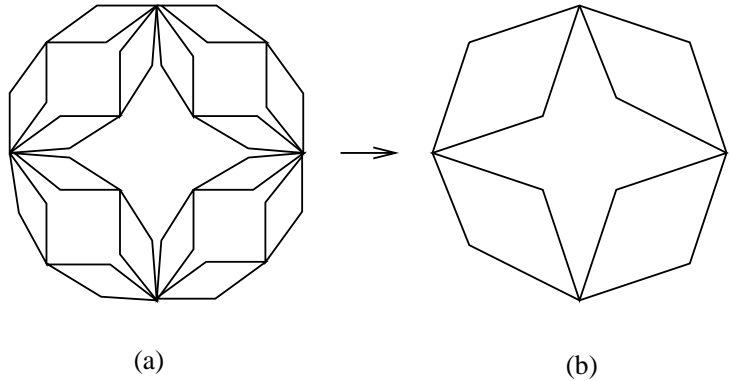,height=5cm}}
\caption{The hierarchical diamond
used in the Migdal-Kadanoff renormalization.
(a) A hierarchical lattice with 3 generations.
(b) A hierarchical lattice with 2 generations.
One RG transformation consists in decimating
the sites at the deepest into the lattice,
{\sl i. e.} for instance transforming the lattice
(a) into the lattice (b).
}
\label{fig:diamond}
\end{figure}

\subsection{Motivation of the hierarchical lattice
study}

In low doping experiments in CuGeO$_3$, Manabe {\sl et al.}
have measured the doping dependence of the
N\'eel temperature, and found that the experimental
data in the range $x > 0.1 \%$ could be well fitted by
the behavior $T_N \sim A \exp{(-B/x)}$~\cite{Manabe98}.
This suggests that there is no critical concentration
associated to antiferromagnetism. As already
pointed out in Refs.~\cite{Fabrizio99,Melin00},
the model Eqs.~\ref{eq:H},~\ref{eq:J}
shows sings of compatibility with these
experiments. The
main unsolved question raised by the experiments
by Manabe {\sl et al.}~\cite{Manabe98} 
is to determine whether the maximum in the
susceptibility
is really the signature of an antiferromagnetic
phase transition with a diverging staggered correlation length.
In fact, the susceptibility experiments give no information
about the existence~/ absence of a diverging correlation
length, and there are are no neutron experiments
with a concentration of impurities of order $\simeq 0.1 \%$.
On the theoretical side,
we used previously several approaches
to describe the nature of the antiferromagnetic phase
of the model Eqs.~\ref{eq:H},~\ref{eq:J}:
a Stoner model~\cite{Fabrizio99},
a decimation method, a
cluster RG~\cite{Melin00}, and a Bethe-Peierls
solution of the classical model~\cite{Melin00}.
It appears that different treatments of the model
have lead to different answers. For instance,
there is a well defined transition in the
Stoner criterion and the Bethe-Peierls treatments~\cite{Fabrizio99,Melin00}.
On the contrary, there is no transition in the decimation
method where the Hamiltonian
Eqs.~\ref{eq:H},~\ref{eq:J} is mapped onto a percolation
model~\cite{Fabrizio99}. A possible approach to this problem
would be to generalize the work in Ref.~\cite{Lima}:
instead of considering the model Eqs.~\ref{eq:H},~\ref{eq:J}
with infinite range exponential interactions,
it is possible to approximate the problem by considering
the Voronoi lattice model with exponential interactions,
in which each lattice site has a finite number of neighbors.
This model is well suited for carrying out numerical
simulations, and avoids the difficulty that the initial
model Eqs.~\ref{eq:H},~\ref{eq:J} has infinite range interactions.

Here, we would like to follow a different route
and replace the original Hamiltonian
Eqs.~\ref{eq:H},~\ref{eq:J} by a simplified one.
This is done by noticing that the essential feature
of the Hamiltonian Eqs.~\ref{eq:H},~\ref{eq:J} is that
the exchange between two spins at a random position
is distributed according to
\begin{equation}
\label{eq:ex-dist}
{\cal P}(J) = \frac{ x \xi}{\Delta}
\left( \frac{J}{\Delta} \right)^{x \xi -1}
,
\end{equation}
where we used the spacing distribution
$P(l) = x \exp{(-xl)}$ and the exchange
$J(l) = \Delta \exp{(-l/\xi)}$. We replace
the original model Eqs.~\ref{eq:H},~\ref{eq:J}
by another model in which the spins are on the
sites of a regular square lattice, and have a random
nearest neighbor exchange in the
distribution~(\ref{eq:ex-dist}). Because the square lattice
model with the exchange distribution Eq.~\ref{eq:ex-dist}
and the original model Eqs.~\ref{eq:H},~\ref{eq:J}
are controlled by the same type of
disorder, it is natural to conjecture that
the two models have an identical physics.
One way to study the square lattice model with the exchange
distribution Eq.~\ref{eq:ex-dist}
would be to perform
large scale Monte Carlo simulations. There is however
a more direct way to handle the model, which consists in
replacing the square lattice by a
recursive
hierarchical lattice (see Fig.~\ref{fig:diamond}),
where the Migdal Kadanoff RG equations can be 
obtained in an exact form.
We will show that
a non trivial physics is going on in the hierarchical
lattice model, which is an indication there is
also a non trivial physics in the square lattice
model with the exchange distribution Eq.~\ref{eq:ex-dist}.

\begin{figure}
\centerline{\psfig{file=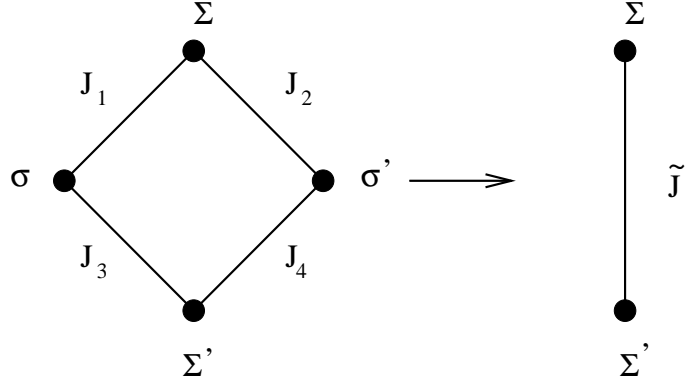,height=5cm}}
\caption{The notations used in the renormalization.
}
\label{fig:renor}
\end{figure}

\subsection{Renormalization group equations}
\label{sec:RGeq}
Let us derive the RG
equations of the Ising model with the exchange distribution
Eq.~\ref{eq:ex-dist} on a hierarchical lattice.
The partition function associated to the exchange configuration
on Fig.~\ref{fig:renor}-(a) reads
\begin{equation}
\label{eq:Z-RG}
Z(\Sigma,\Sigma') = \sum_{\sigma,\sigma'}
\exp{ \beta\left[ J_1 \Sigma \sigma
+ J_2 \Sigma \sigma' + J_3 \Sigma'
\sigma + J_4 \Sigma' \sigma' \right]}
,
\end{equation}
and we impose that Eq.~\ref{eq:Z-RG} be identical
to the partition function associated to the
exchange configuration on Fig.~\ref{fig:renor}-(b),
up to a proportionality factor:
$$
Z(\Sigma,\Sigma') = {\cal N} \exp{(\beta
\tilde{J} \Sigma \Sigma')}
.
$$
Using the relation
$$
\tilde{J} = \frac{1}{2 \beta} \ln{ \left[
\frac{ Z(+,+)}{Z(+,-)} \right]}
,
$$
we find
$\tilde{J} =
\tilde{J}_{1-3} + \tilde{J}_{2-4}$, where
\begin{equation}
\label{eq:J13}
\tilde{J}_{1-3} = \frac{1}{2 \beta}
\ln{ \left[ \frac{ \cosh{(\beta (J_1 + J_3))}
}
{\cosh{(\beta (J_1 - J_3))}} \right]}
\mbox{ , }
\tilde{J}_{2-4} = \frac{1}{2 \beta}
\ln{ \left[ \frac{ \cosh{(\beta (J_2 + J_4))}
}
{\cosh{(\beta (J_2- J_4))}} \right]}
.
\end{equation}
Now that we have determined the renormalization
of the exchanges, we iterate
the exchange distribution.
Noting $P(J_1)$ ... $P(J_4)$ the distribution
of the exchanges $J_1$ ... $J_4$, and $P(\tilde{J})$
the distribution of the renormalized exchange $\tilde{J}$, we
have
\begin{equation}
\label{eq:iter}
\tilde{P}(\tilde{J}) = 
\int d J_1 d J_2 d J_3 d J_4
P(J_1) P(J_2) P(J_3) P(J_4)
\delta \left[ \tilde{J} - \tilde{J}_{1-3}
- \tilde{J}_{2-4} \right]
.
\end{equation}
It will be useful to change
variables to the bond lengths
$l = \xi \ln{(\Delta/J)}$ and iterate
the bond length distribution $p(l)$
instead of the exchange distribution $P(J)$.

To calculate numerically the iteration of the exchange
distribution Eq.~\ref{eq:iter},
we use a discrete bond length $l=1,...,N$, and we
introduce an upper cut-off for the bond length.
This is valid if we can check that the RG flow
does not depend on $N$.

\subsection{Analysis of the RG flow}

\begin{figure}
\centerline{\psfig{file=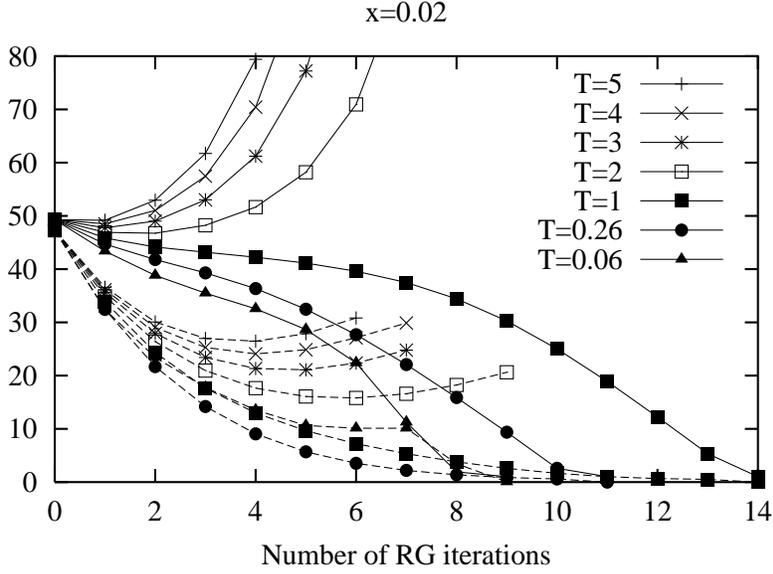,height=8cm}}
\caption{RG flow of the classical model relevant
for the antiferromagnetic transition in CuGeO$_3$.
The parameters are $\Delta=44.7$~K
and $\xi=10$, and we used a cut-off $N=300$.
The doping concentration is $x=0.02$. The solid
line represents the evolution of the average
bond length $\LL l \RR$. The doted lines
represent the evolution of the width of the bond length
distribution $\sqrt{\LL (l - \LL l \RR)^2 \RR}$.
The RG trajectories with
$T=5$ ($+$), $T=4$ ($\times$), $T=3$ (*) and
$T=2$ ($\Box$) flow into
the paramagnetic phase (the bond length renormalizes
to infinity).
The trajectories with $T=1$ ($\blacksquare$),
$T=0.26$ ($\bullet$), $T=0.06$ ($\blacktriangle$)
flow into the ferromagnetic phase (the bond
length renormalizes to zero).
}
\label{fig:RG-flow}
\end{figure}

The RG flow of the model is shown
on Fig.~\ref{fig:RG-flow}. We have shown
on this figure the evolution of the average
bond length $\LL l \RR$ and the width of the
bond length distribution $\sqrt{\LL (l - \LL l
\RR)^2 \RR}$. At low temperature, the average bond
length renormalizes to zero (ordered phase)
while at high temperature it renormalizes to
infinity (paramagnetic phase). Therefore, there
is a well-defined transition in the model. We have
checked that the transition temperature is independent
on the cut-off $N$ used to iterate the bond length
distribution. The existence of a thermodynamic transition
could have been anticipated on the basis of the
simplest possible approximation of the RG flow
(see Appendix~\ref{sec:proj}). What is less obvious is that,
after a transient in the first RG iterations,
the width of the exchange distribution becomes
much smaller than the average exchange: in spite
of a broad initial exchange distribution
in which $\LL l \RR = \sqrt{ \LL ( l - \LL l \RR)^2 \RR}$,
the system renormalizes to an almost
disorder-free exchange distribution in which
$\sqrt{ \LL ( l - \LL l \RR)^2 \RR} \ll \LL l \RR$.
Therefore, at large scale, the spin system looks
ordered while inhomogeneities are visible only
at small scale. This type of behavior may explain
why there is a pronounced maximum in the temperature
dependence of the susceptibility even at very low
doping~\cite{Manabe98}.

\section{Nature of the quantum disordered region}
\label{sec:nature-qu}

Now, we would like to investigate the behavior of the
model in the quantum disordered region, and compare
it to experimental data on the Haldane gap compound
Y$_2$BaNiO$_5$. More specifically,
we would like to determine whether the behavior
of the model is compatible with the susceptibility
experiments by Payen {\sl et al.}~\cite{Payen00}, who
reported that the susceptibility of Y$_2$BaNiO$_5$
has a power-law temperature dependence.
In our opinion, it is an important question to
determine which ingredients should be incorporated
in the theoretical model to describe the existing
experiments. A first strategy, followed by
Batista {\sl et al.}~\cite{Batista} is to look
for the ``most realistic possible'' model. The first
step in this approach is to consider that the 
relevant Hamiltonian for Y$_2$BaNiO$_5$ takes the form
\begin{equation}
\label{eq:H-Bat}
{\cal H} = \sum_i \left\{ J {\bf S}_i. {\bf S}_{i+1}
+ D (S_i^z)^2 + E \left[ (S_i^x)^2 - (S_i^y)^2 
\right] \right\}
.
\end{equation}
The anisotropy parameters in Eq.~\ref{eq:H-Bat}
have been
determined by fitting inelastic
neutron scattering experiments~\cite{fit-spin1-a,fit-spin1-b}.
Next, a density matrix renormalization
group (DMRG) method has been used to treat the Hamiltonian
Eq.~\ref{eq:H-Bat} in the presence of magnetic
impurities. The authors
of Ref.~\cite{Batista} are then able to reproduce
specific heat experiments,
and also arrive to an agreement with susceptibility
experiments~\cite{Payen00}.
The main objection that one might be tempted
to formulate is that the Hamiltonian
of the spin-$1$ chain relevant for Y$_2$BaNiO$_5$ contains
already three adjustable parameters, and that
two more additional gyromagnetic factors have been
added to describe the susceptibility experiments~\cite{Payen00}.
Here, it is proposed that the power law temperature
dependence of the susceptibility is a generic feature of
the quantum disordered region of the phase diagram,
that can be explained in a model with only a
minimal number of ingredients.

\begin{figure}
\centerline{\psfig{file=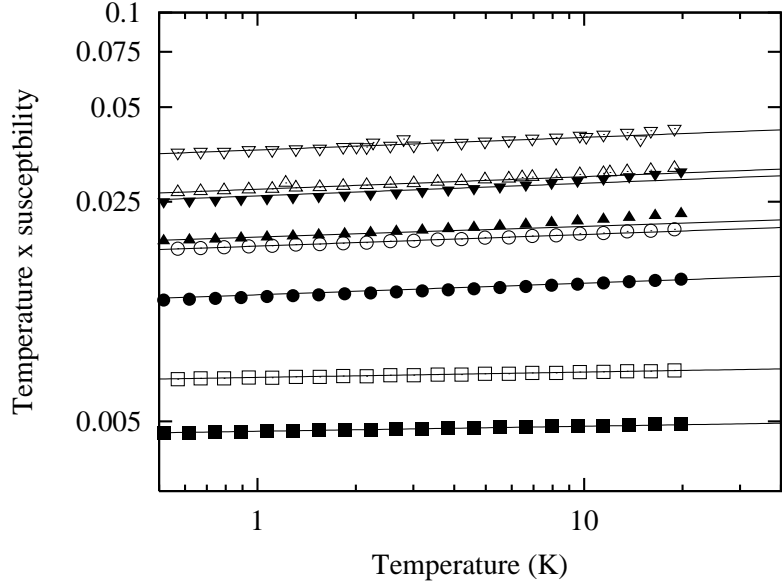,height=8cm}}
\caption{Log-log plot of the temperature dependence of
$T \chi(T)$ for the two-spin model and the cluster
RG of the full chain. We used $\Delta = 100$~K, $\xi=6$.
For the two-spin model,
we used $x=0.01$ ($\Box$),
$x=0.03$ ($\odot$), $x=0.05$ ($\Delta$), $x=0.07$
($\nabla$). The same symbols filled in black have been used for
the cluster RG calculation. The product $T \chi(T)$ has
been fitted to a power-law dependence $T \chi(T) \sim
T^\alpha$, with $\alpha=0.017$ ($x=0.01$),
$\alpha=0.037$ ($x=0.03$), $\alpha=0.04$ ($x=0.05$),
$\alpha=0.04$ ($x=0.07$).
}
\label{fig:qu}
\end{figure}

\begin{figure}
\centerline{\psfig{file=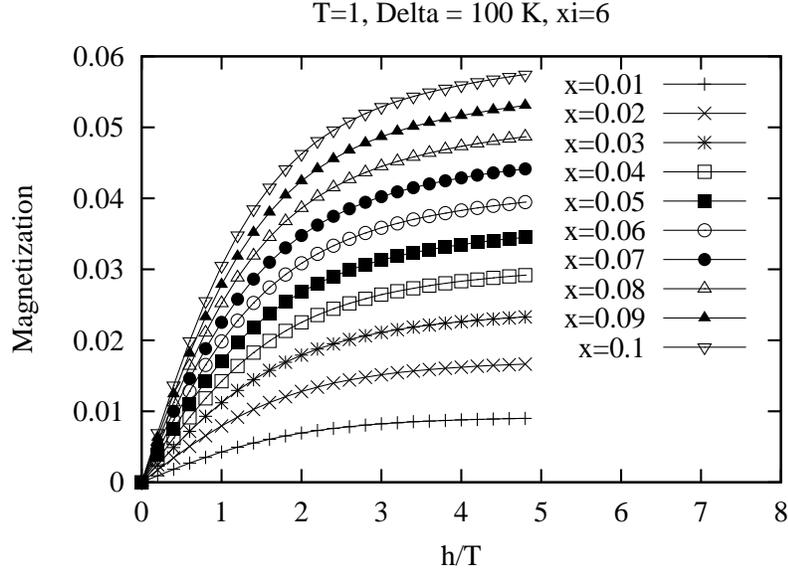,height=8cm}}
\caption{Magnetization of the
two-spin quantum model, with the parameters relevant
for Y$_2$BaNiO$_5$: $\Delta=100$~K, $\xi=6$. 
}
\label{fig:Mh}
\end{figure}

\subsection{Two-spin model}
\label{sec:2spin}

As we already explained in the Introduction,
in the low energy model relevant to describe the doped
Haldane gap compound, each paramagnetic
impurity generates a unit of two ``edge'' spin-$1/2$
moments (see Fig.~\ref{fig:schema}). The two spin-$1/2$
moments in the same unit are coupled by a ferromagnetic
exchange $J_2$ the magnitude of which is of order
of interchain interactions~\cite{Fabrizio-spin1}.
As shown on Fig.~\ref{fig:schema},
there is a staggered exchange 
$(-)^l \exp{(-l/\xi)}$ coupling two edge moments
at a distance $l$. 

Let us first consider the simplest model
in which the impurities are assumed to cut the
chain into finite segments: $J_2=0$. This is a valid
model if the temperature is larger than $J_2$, or
equivalently, than the strength of
interchain interactions.
Consider
two edge spins at distance $l$, coupled with a
Heisenberg Hamiltonian
${\cal H} = J(l) {\bf S}_1 . {\bf S}_2 - h
({\bf S}_1^z + {\bf S}_2^z) $, and
the exchange $J(l) = \Delta (-)^l \exp{(-l/\xi)}$,
with ${\cal P}(l_+)=x \exp{(-x l_+)}$ the distribution
of even length segments and
${\cal P}(l_-)=x \exp{(-x l_-)}$ the distribution
of odd length segments.
It is easy to calculate
the average magnetization in a magnetic field $h$:
\begin{equation}
\label{eq:MM1}
\langle \langle M(h,T) \rangle \rangle
=\frac{1}{2}x \int_0^\Delta {\cal P}(J) \left[ M_+(h,T,J)+M_-(h,T,J) \right] dJ
,
\end{equation}
with ${\cal P}(|J|) = (x \xi / \Delta) (|J|/\Delta)^{-1+x \xi}$
the exchange distribution, and
\begin{equation}
\label{eq:MM2}
M_+(h,T,J)= \frac{e^{\beta h} - e^{- \beta h}}
{ e^{\beta h} + e^{- \beta  h} +1 + e^{\beta J}}
\mbox{ , }
M_-(h,T,J)= \frac{e^{\beta h} - e^{- \beta h}}
{ e^{\beta h} + e^{- \beta  h} +1 + e^{-\beta J}}
\end{equation}
the magnetization of the spins coupled by an antiferromagnetic
or ferromagnetic exchange $J$ at a finite temperature
$T = 1 / \beta$.
To calculate the magnetization Eqs.~\ref{eq:MM1},~\ref{eq:MM2},
it is convenient to integrate by parts:
\begin{equation}
\label{eq:M1}
\langle \langle M(h,T)
\rangle \rangle = x A +
x \left( \frac{T}{\Delta} \right)^{x \xi}
\int_0^{\Delta/T} u^{x \xi} f(u,h/T)
,
\end{equation}
with
\begin{eqnarray}
\label{eq:M2}
A  &=& \frac{1}{2} \left[ \frac{ e^{ \beta h}
- e^{ - \beta h}}
{ e^{ \beta h}
+ e^{ - \beta h} + 1 +e^{\Delta/T}}
+ \frac{ e^{ \beta h}
- e^{ - \beta h}}
{ e^{ \beta h}
+ e^{ - \beta h} + 1 +e^{-\Delta/T}} \right]
\\
f(u,\frac{h}{T})&=& -  \frac{1}{2}
\frac{d}{du} \left[
\frac{ e^{\beta h} - e^{-\beta h}}
{ e^{\beta h} + e^{-\beta h} + 1+ e^{u}}
+ \frac{ e^{\beta h} - e^{-\beta h}}
{ e^{\beta h} + e^{-\beta h} + 1+ e^{-u}}
\right]
.
\label{eq:M3}
\end{eqnarray}
The resulting susceptibility is shown on Fig.~\ref{fig:qu},
where it is visible that the product $T \chi(T)\sim T^\alpha$
has a power-law behavior at low temperature, like
what has been observed in experiments on
Y$_2$BaNiO$_5$~\cite{Payen00}.
The magnetization in an applied magnetic field
is shown on Fig.~\ref{fig:Mh}, and
is close to the experimental observation~\cite{Payen00}.
This shows that this model with only
two spins contains much of the physics as far as the temperature
is above the interchain coupling $J_{\perp}$. 

\subsection{Cluster RG}
\label{sec:clusterRG}

\subsubsection{Quantum disordered phase I}
Now, let us 
consider the cluster RG of the model with a finite
$J_2$, and first consider the behavior of the model
when the temperature is larger than $J_2$. I refer
the reader to Refs.~\cite{Dasgupta,Westerberg,Melin00}
for an explanation of the
method, and just present here the results. First,
if the temperature is above $J_2$, we find that the exponent
of the power-law Curie constant $T \chi(T) \sim T^{\alpha}$
is identical to the one of the two-spin model (see Fig~\ref{fig:qu}).
This is  not a surprise because above $J_2$ 
the edge moments in the same unit remain uncoupled
and the cluster RG contains the same physics
as the two-spin model. Moreover, the two-spin
model is an exact treatment while the cluster
RG is approximate. The agreement between the two
methods shows the validity of the cluster RG method.

\begin{figure}
\centerline{\psfig{file=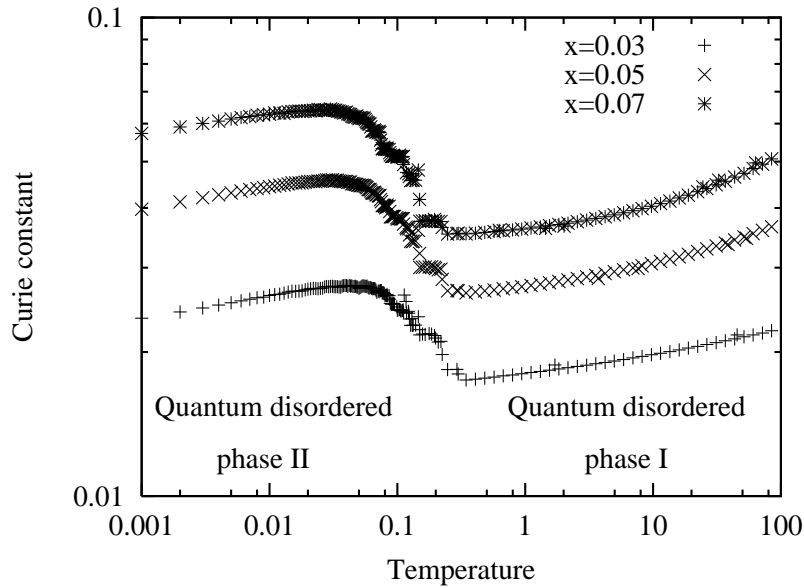,height=8cm}}
\caption{Temperature dependence of the Curie
constant, with the parameters relevant
for Y$_2$BaNiO$_5$: $\Delta=100$~K, $\xi=6$,
and $x=0.03$, $x=0.05$, and $x=0.07$.
We renormalized a chain with $10000$ sites and
averaged over $100$ realizations of disorder.
We used $J_2=0.3$~K.
}
\label{fig:Curie}
\end{figure}

\subsubsection{Quantum disordered phase II}

Now, I consider the cluster RG of a single chain
at a temperature smaller than $J_2\sim J_{\perp}$.
The temperature dependence of the Curie constant
is shown on Fig.~\ref{fig:Curie}. It is visible that
the Curie constant increases strongly with
decreasing the temperature below $J_2$. This is because when
$T \sim J_2$, the survival spin-$1/2$ moments in the same unit
are frozen into spin-$1$ moments. This phenomenon is not
quantum in nature because it occurs also in the classical
disordered model, which is analyzed in details below.

Once the freezing into spin-1 units has been done at the
energy scale $J_2$, the resulting effective model
is again the one of spin-$1$ objects with random
exchanges, being ferromagnetic or antiferromagnetic.
This results in the low temperature
quantum disordered phase II on Fig.~\ref{fig:Curie}.

In the presence of a finite interchain coupling, one still
expects the appearance of two types of quantum
disordered regions. It is also expected that the
quantum disordered phase II on Fig.~\ref{fig:Curie}
is a 3D random singlet state, with singlet formation
between spins belonging to different chains. The
cross-over to a 3D regime is however
not expected to change the shape of the
temperature dependence of the susceptibility
(see Fig.~\ref{fig:Curie}).

\subsection{Classical disordered model}
\label{sec:collective}
Now, let us  determine to what extend
the physics of the classical magnet resembles the physics
of the quantum magnet.
The Ising chain Hamiltonian reads
$$
{\cal H} = \sum_{i} J_{i,i+1} \sigma_i
\sigma_{i+1}
,
$$
with the Ising variables $\sigma_i$
corresponding to the consecutive edge moments, and
$J_{i,i+1}$ determined according to the rules
on Fig.~\ref{fig:schema}. The even bonds correspond
to a ferromagnetic exchange $-J_2$ and
the odd bonds correspond to an exchange $J(l) = \Delta
(-)^l \exp{[-l/\xi]}$, with the spacing $l$
drawn in the Poisson distribution
$P(l)= x \exp{(-xl)}$. 

As shown in Appendix~\ref{app:spin1},
the Curie constant crosses over
from $2x$ at high temperature
to $4x$ at low temperature, when the temperature
decreases below $J_2$. Therefore, in the classical
model, the product
$T \chi(T)$ increases monotonically with a decreasing temperature,
while the opposite is observed experimentally in
Y$_2$BaNiO$_5$~\cite{Payen00}. This is not unexpected
because the classical model cannot describe the
gapless Haldane phases. Nevertheless, the increase
in the susceptibility below $J_2$ is very reminiscent
of the behavior of the quantum model in the same
temperature range. This is because the freezing
of the edge moments of the same unit in a ferromagnetic
alignment occurs both in the classical and quantum
models. 

\begin{figure}
\centerline{\psfig{file=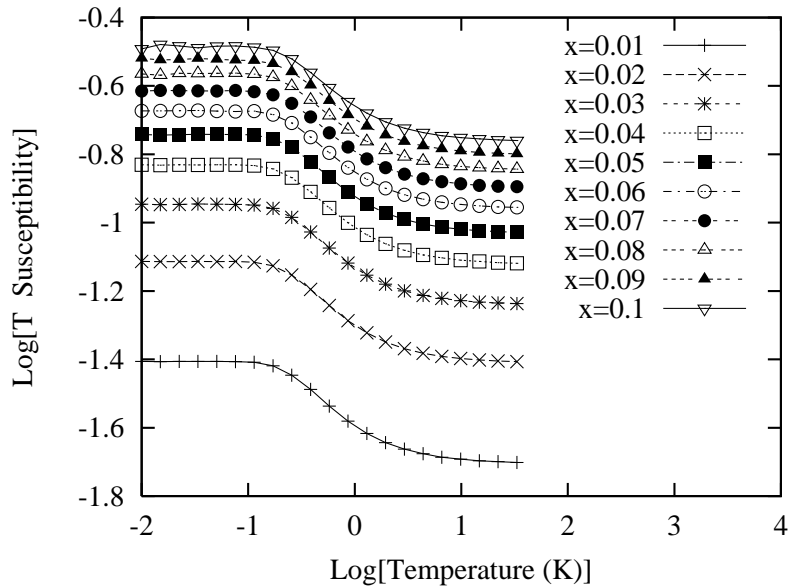,height=8cm}}
\caption{Temperature dependence of the Curie
constant $T \chi(T)$ in the classical analog of the
doped spin-$1$ chain.
The parameters are $\xi=6$, $\Delta=100$~K,
$J_2 = 0.3$~K, relevant for Y$_2$BaNiO$_5$. The Curie constant
crosses over from $2x$ at high temperature to
$4x$ at low temperature.
}
\label{fig:chi-T}
\end{figure}

\begin{figure}
\centerline{\psfig{file=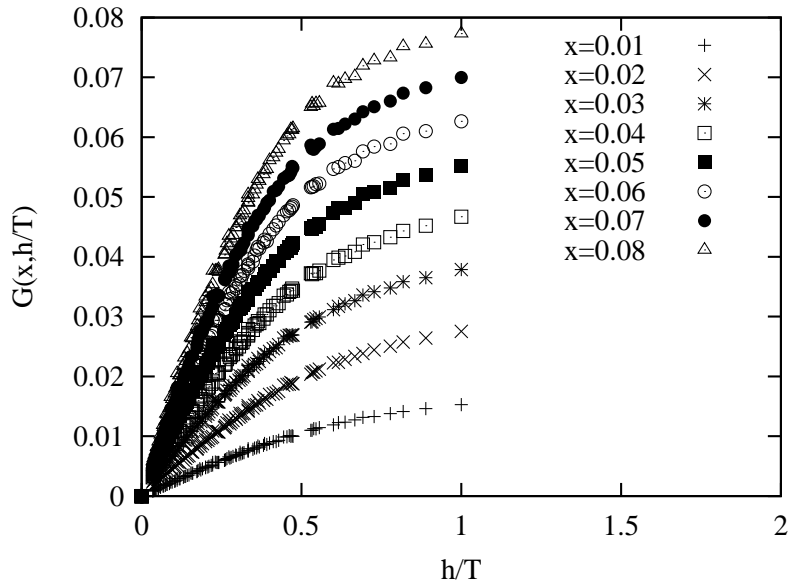,height=8cm}}
\caption{Scaling of the magnetization
$M(h,T) = T^{-\gamma} G(x,h/T)$ in the classical
analog of the doped spin-$1$ chain.
We have
shown $G(x,h/T)$ as a function of $h/T$.
Each curve corresponds to values of
$T$ varying from $0.9$ to $3$ and $h$
varying from $0$ to $1$. We used the parameters
$\xi=6$, $\Delta=100$~K relevant for Y$_2$BaNiO$_5$.
We used $\gamma \simeq 0.05$.
}
\label{fig:master}
\end{figure}

The scaling function of the magnetization can be calculated
easily by iterating numerically the magnetization
distribution Eq.~\ref{eq:PM} and using the relation
$$
{\cal P}(M,h) = \frac{ {\cal P}(M,h=0) e^{\beta h M}}
{ \sum_{M'} {\cal P}(M',h=0) e^{\beta h M'}}
$$
to obtain the magnetization distribution in a finite
magnetic field. The magnetization takes the
form $M(h,T) = T^{-\gamma} G(x,h/T)$ in a given
temperature range where the susceptibility
can be approximated by $\chi(T) \sim T^{-1-\gamma}$.
The scaling function is shown on Fig.~\ref{fig:master},
where it is visible that it has qualitatively the
correct behavior in spite of the exponent
$\gamma$ having the wrong sign compared to experiments.
Therefore, the shape
of the scaling function of the magnetization
does not appear to be a crucial test to the model.

\section{Conclusions}
\label{sec:conclusion}

To conclude, the present work was intented to 
describe doping a spin-Peierls and a Haldane gap
state in a unified framework.
We have first shown how the relevant
energy scales in the problem could be calculated from
the analysis of a two-spin model, which allowed to
discuss the phases of the model as a function of
the doping concentration and interchain interactions.
In the relevant temperature window, there are two
distinct regions depending on the doping concentration
and interchain interactions: (i) an antiferromagnetic region;
(ii) a quantum disordered region. remarkably, this
type of phase diagram compares well with the known
behavior of the spin-Peierls compound CuGeO$_3$ and the two
Nickel oxides PbNi$_2$V$_2$O$_8$ and Y$_2$BaNiO$_5$.

Next, we used our approach to investigate in more details
the two possible phases of the model. We have shown that
the physics in the antiferromagnetic region of the
phase diagram is {\sl classical in nature} and
therefore we were lead to study the corresponding
Ising model. We have replaced the original Hamiltonian
Eqs.~\ref{eq:H},~\ref{eq:J} by another Hamiltonian
having the same features, and presented the solution
of the latter Hamiltonian on a hierarchical lattice structure.
Interestingly, we find that the renormalized problem
is non disordered. This is in agreement with the presence
of a well-defined cusp associated to antiferromagnetism
in the susceptibility~\cite{Manabe98}.

In the ``quantum disordered'' region of the phase diagram,
the physics is strongly controlled by quantum fluctuations.
Already in a model with two spins only does the susceptibility
have a power-law temperature dependence, very similar to
the experimental observation~\cite{Payen00}. We have suggested
that there is no need to introduce many coupling constants
to reach an agreement between the model and experiments. 
Going beyond the level of a two-spin model, we have found
the existence of another quantum disordered phase at low
temperature. We have analyzed the behavior of the classical
model and shown it contains already a physics relevant
to the quantum model.

\appendix

\section{Projection of the RG flow on a
trial exchange distribution}
\label{sec:proj}
We would like to present the simplest possible
approximation of the RG equations obtained in
section~\ref{sec:RGeq} in which
we project the RG flow 
onto the single parameter distribution
$p_{n}(l) = \delta(l-L_n)$. The initial distribution
is obtained {\sl via} the relation
$\int p_0(l) dl = \int x \exp{(-x l)} dl$, with
$p_0(l) = \delta( l - L_0)$. This leads to
\begin{equation}
\label{eq:L0}
L_0 = \xi \ln{
\left( \frac{1 + x \xi}{x \xi}\right)}
.
\end{equation}
Next, we start from the
distribution $p_n(l)$, make one RG
transformation Eq.~\ref{eq:iter}, and impose
that the iterated distribution has the same
first moment as $p_{n+1}(L)$, from what
we can determine the parameter $L_{n+1}$:
\begin{equation}
\label{eq:Ln+1}
L_{n+1} =  \xi \ln{ \frac{ \beta \Delta}
{ \ln { \cosh{ \left[ 2 \beta \Delta
\exp{(-L_n/\xi)} \right]}}}}
.
\end{equation}
Since we want to discuss the stability of the
paramagnetic phase, we consider Eq.~\ref{eq:Ln+1}
in the limit of a large bond length:
$L_{n+1} \simeq - \xi \ln{(2 \beta \Delta)} + 2 L_n$.
In this limit, we find
$
L_n = \xi \ln{(2 \beta \Delta)}
+ \left( L_0 - \xi \ln{(2 \beta \Delta)} \right)
2^n
.
$
Using Eq.~\ref{eq:L0}, we obtain
a phase transition at the temperature
$$
T_c = \frac{2 \Delta x \xi}{1 + x \xi}
.
$$
The transition
temperature is in agreement with what has been found
in previous works with different methods (the Stoner
criterion~\cite{Fabrizio99} and the Bethe-Peierls
method~\cite{Melin00}). As we show in the body of the article,
the renormalized exchange distribution can be well
approximated by the distribution $p_{n}(l) = \delta(l-L_n)$
in the sense that the problem renormalizes to a non disordered
one.

\section{Solution of the classical analog of the
doped spin-$1$ chain}
\label{app:spin1}
We consider a finite chain
with $N$ edge moments in which the end spin at site $N$
is frozen in the direction $+$, and note
${\cal P}_N^+(M)$
the corresponding magnetization distribution.
We note $x_i = \exp{(\beta J_i)}/
[ \exp{(\beta J_i)} + \exp{(-\beta J_i)}]$
the probability to find the spins
$\sigma_i$ and $\sigma_{i+1}$ in an
antiparallel alignment. We have
\begin{equation}
\label{eq:PM}
{\cal P}_{N+1}^+(M) = (1-x_N) {\cal P}_{N}^+(M-1)
+ x_N {\cal P}_N^{-}(M-1)
.
\end{equation}
Using the relation ${\cal P}_N^+(M) = {\cal P}_N^-(-M)$,
we get
\begin{eqnarray}
\langle M \rangle_{N+1}^+ &=& (1 - 2 x_N )
\langle M \rangle_{N}^+ +1 \\
\langle M^2 \rangle_{N+1} &=&
\langle M^2 \rangle_{N+1} + 2 (1- 2 x_N)
\langle M \rangle_N^+ +1
.
\end{eqnarray}
These relations can be solved analytically.
For this purpose, let us separate the even bonds
coupled by the ferromagnetic exchange
$-J_2$ and note $x_F = e^{-\beta J_2}/
[e^{\beta J_2} + e^{-\beta J_2}]$.
The odd bonds are ferromagnetic or antiferromagnetic
and we are lead to define
$$
y = \sum_{l=1}^{+ \infty} {\cal P}(l)
\frac{ e^{-\beta J(l)} }{ e^{\beta J(l)}
+ e^{-\beta J(l)}}
.
$$
The magnetization of a chain with an even number of sites
$N=2p$
is found to be
\begin{equation}
\label{eq:M-even}
\langle M \rangle_{2p}^+ = \frac{2 (1-x_F)}{1-X}
\left[ 1 - X^p \right]
,
\end{equation}
with $X=(1-2 x_F) (1-2 y)$. The correlation length
is $\xi_T = -2 / \ln{X}$.
With an odd number of sites $N=2p+1$, the magnetization is
\begin{equation}
\label{eq:M-odd}
\langle M \rangle_{2p+1}^+ = 
\frac{2(1-y)}{1-X} - \frac{X+1-2y}{1-X} X^p
.
\end{equation}
The expressions of the first moment
Eqs.~\ref{eq:M-even},~\ref{eq:M-odd} are next used
to solve for the second moment:
$$
\langle M^2 \rangle_{2p} \sim
2p \left\{ 1 + \frac{2(1-2y)(1-x_F)}{1-X}
+ \frac{ 2(1-2 x_F)(1-y)}{1-X} \right\}
,
$$
to leading order in the chain length $2p$.
At high temperature, one has
$y \sim 1/2$, $x_F \sim 1/2$ and 
$X \sim 0$, and therefore a susceptibility
scaling like $\chi \sim 2x/T$. At low temperature,
one has $x_F \sim 0$, $y \sim 1/2$ and $X \sim 0$,
and a susceptibility scaling like $\chi
\sim 4x /T$. Therefore, the Curie constant crosses over
from the high temperature value $2 x$
to $4x$ at low temperature, because the spin-$1/2$
moments in the same unit are frozen ferromagnetically
at a temperature $T \sim J_2$
(see Fig.~\ref{fig:chi-T}).

\end{document}